\documentclass[a4paper,11pt]{article}
\pdfoutput=1 

\usepackage{jinstpub} 
\usepackage{caption}
\usepackage{subcaption}
\usepackage{nicefrac}

\graphicspath{{figures/}}

\title{The ARAGORN front-end -- An FPGA based implementation of a Time-to-Digital Converter}

\author[1]{M. Büchele,\note{Corresponding author.}}
\author{H. Fischer,}
\author{F. Herrmann}
\author{and C. Schaffer}
\affiliation{Physikalisches Institut, Albert-Ludwigs-Universität Freiburg\\
Hermann-Herder-Str. 3, 79104 Freiburg, Germany}

\emailAdd{maximilian.buchele@cern.ch}

\abstract{We present the ARAGORN front-end, a cost-optimized, high-density Time-to-Digital
  Converter platform. Four Xilinx Artix-7 FPGAs implement 384~channels with an average
  time resolution of 165\,ps on a single module. A fifth FPGA acts as data concentrator
  and generic board master. The front-end features a SFP+ transceiver for data output and
  an optional multi-channel optical transceiver slot to interconnect with up to seven
  boards though a star topology. This novel approach makes it possible to read out up to
  eight boards yielding 3072~input channels via a single optical fiber at a bandwidth of
  6.6\,Gb/s.}

\begin{document}
\maketitle
\flushbottom

\section{Introduction}
\label{sec:intro}

\begin{figure}[htbp] \centering 
 \begin{subfigure}[b]{0.37\textwidth}
        \includegraphics[width=\textwidth]{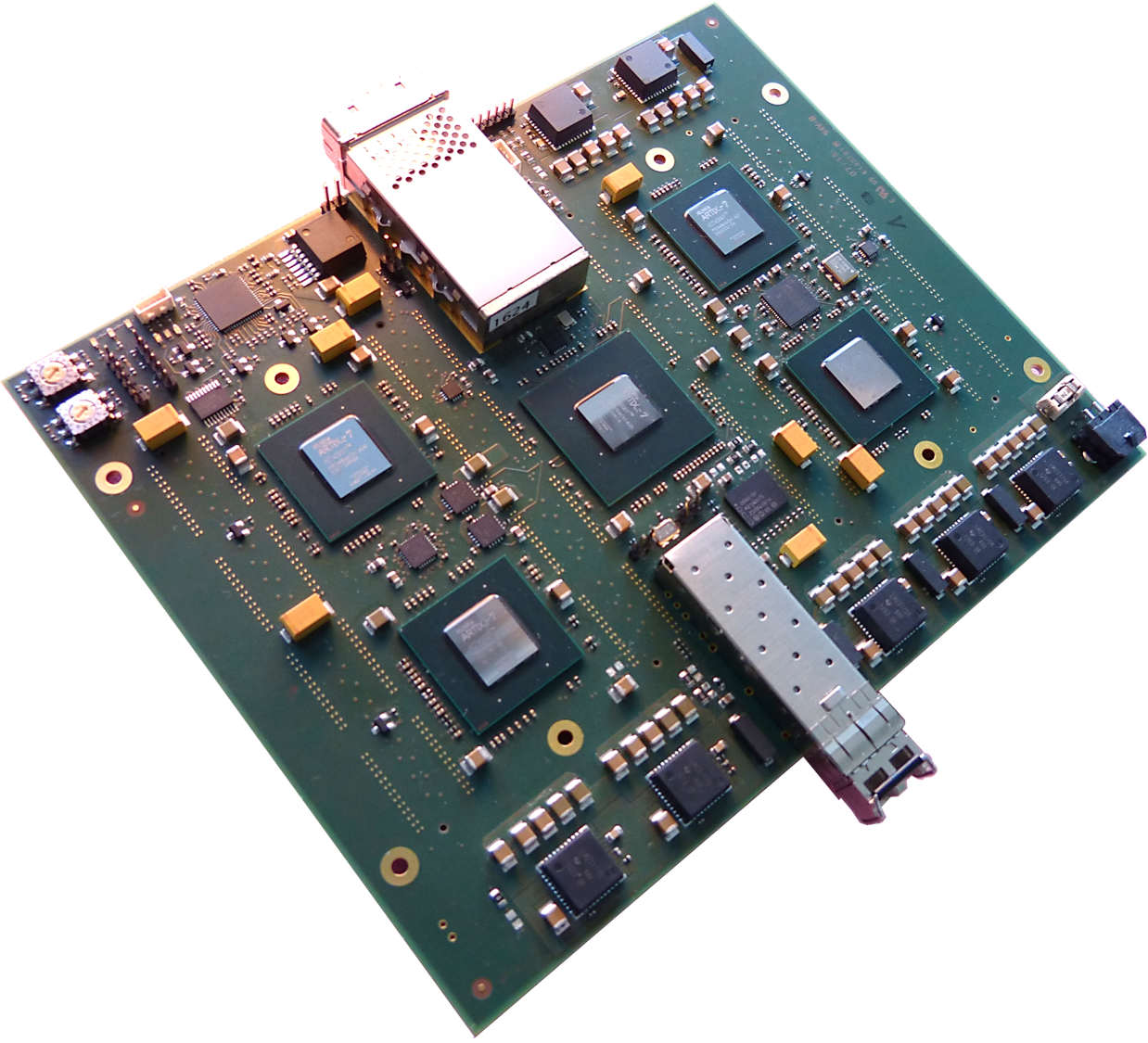}
        \caption{\label{fig:aragorn}}
    \end{subfigure} \hfill
    \begin{subfigure}[b]{0.6\textwidth}
        \includegraphics[width=\textwidth]{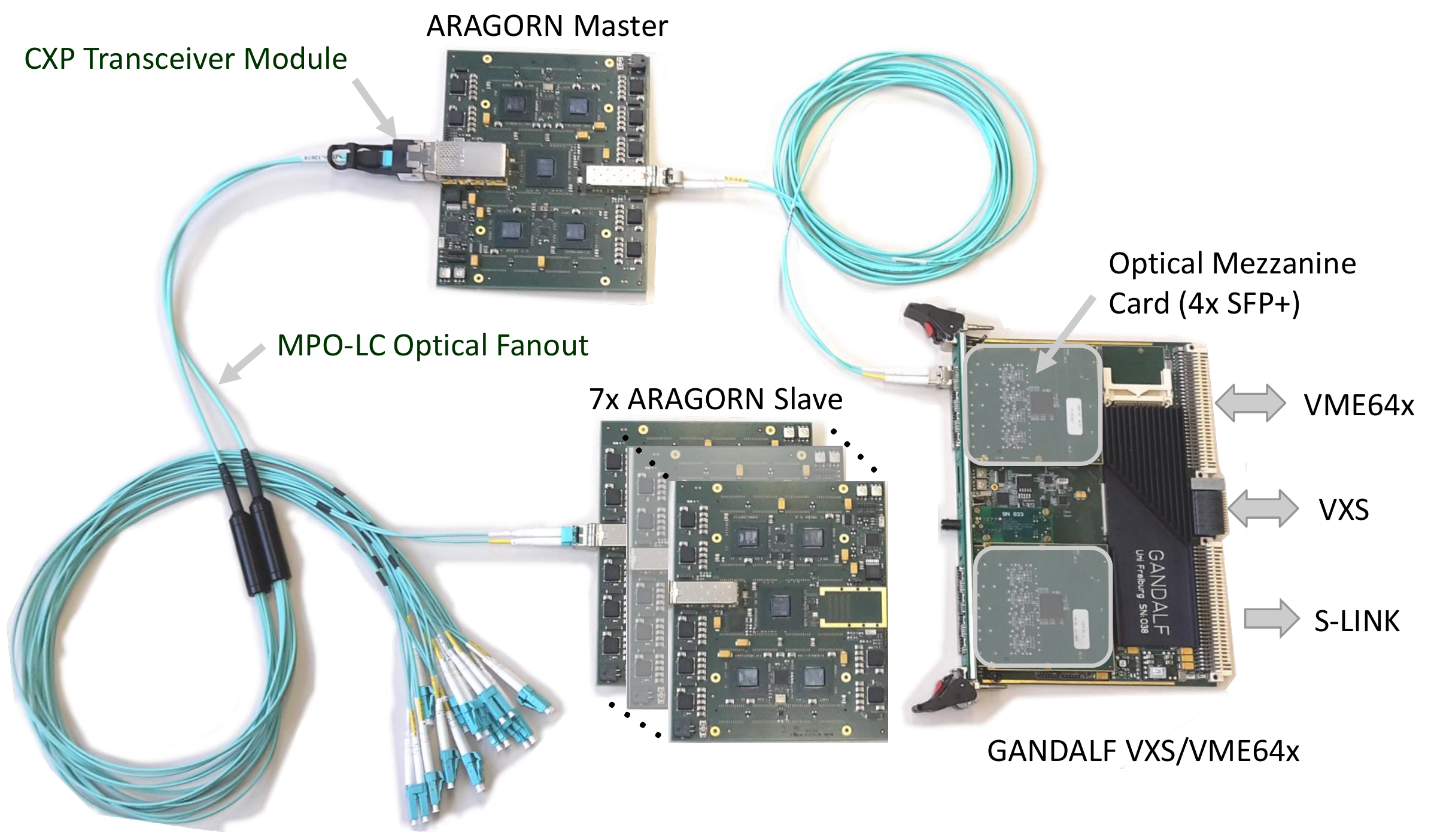}
        \caption{\label{fig:star_topo}}
    \end{subfigure}
    \caption{(\subref{fig:aragorn}) The ARAGORN front-end comprises 4~+~1 Artix-7 FPGAs,
      four of which implement 384 TDC channels. The central FPGA acts as data hub and
      masters the communication with the SFP+ and CXP transceiver
      sockets. (\subref{fig:star_topo}) Star topology interconnection of up to eight
      ARAGORN front-ends leveraging the CXP transceiver module and the
      GANDALF~VXS/VME64x~\cite{gandalf} for data readout.}
\end{figure}

The ARAGORN front-end (figure~\ref{fig:aragorn}) offers high-performance digital readout
capabilities for high-luminosity experimental environments. In order to provide a
competitive design that can be operated in widespread applications, the board employs
4~+~1 Xilinx Artix-7 (xc7a200t) FPGAs. Four of these implement a Time-to-Digital Converter
(TDC) processing 384~channels on a single module, whereas the fifth FPGA acts as data hub
and oversees the communication with auxiliary board components. Implementing the major
board functions with low-cost, fully configurable FPGAs ensures long-term usability of
this hardware platform, adapting to versatile readout tasks within short development time.

The TDC inputs are linked to four high-speed SMD connectors, providing an interface for
extension boards, primarily incorporating the preamplifier and discriminator modules. The
acquired data is transferred via source-synchronous serial links to the central FPGA for
data output. A highlight of this project is the superior optical readout concept ---
leveraging both a SFP+ transceiver slot for data output and an optional CXP transceiver
socket --- to interconnect with up to seven ARAGORN cards as satellites using an optical
fanout cable (figure~\ref{fig:star_topo}). The transceiver modules are in turn attached to
the high-speed transceiver tiles of the central FPGA. Consequently, eight boards thus
3072~TDC channels are read out by a single optical fiber connected to the master board
that hosts the CXP transceiver. The decisive benefit is that the board layout remains
unchanged being operated in master configuration or not. Thanks to its pluggable
implementation, the costly CXP module is only required for the master board application,
thus strongly reducing the operating expenses.
    
\section{TDC architecture}
\label{sec:tdc_arch}

\begin{figure}[h] \centering 
 \begin{subfigure}[b]{0.42\textwidth}
        \includegraphics[width=\textwidth]{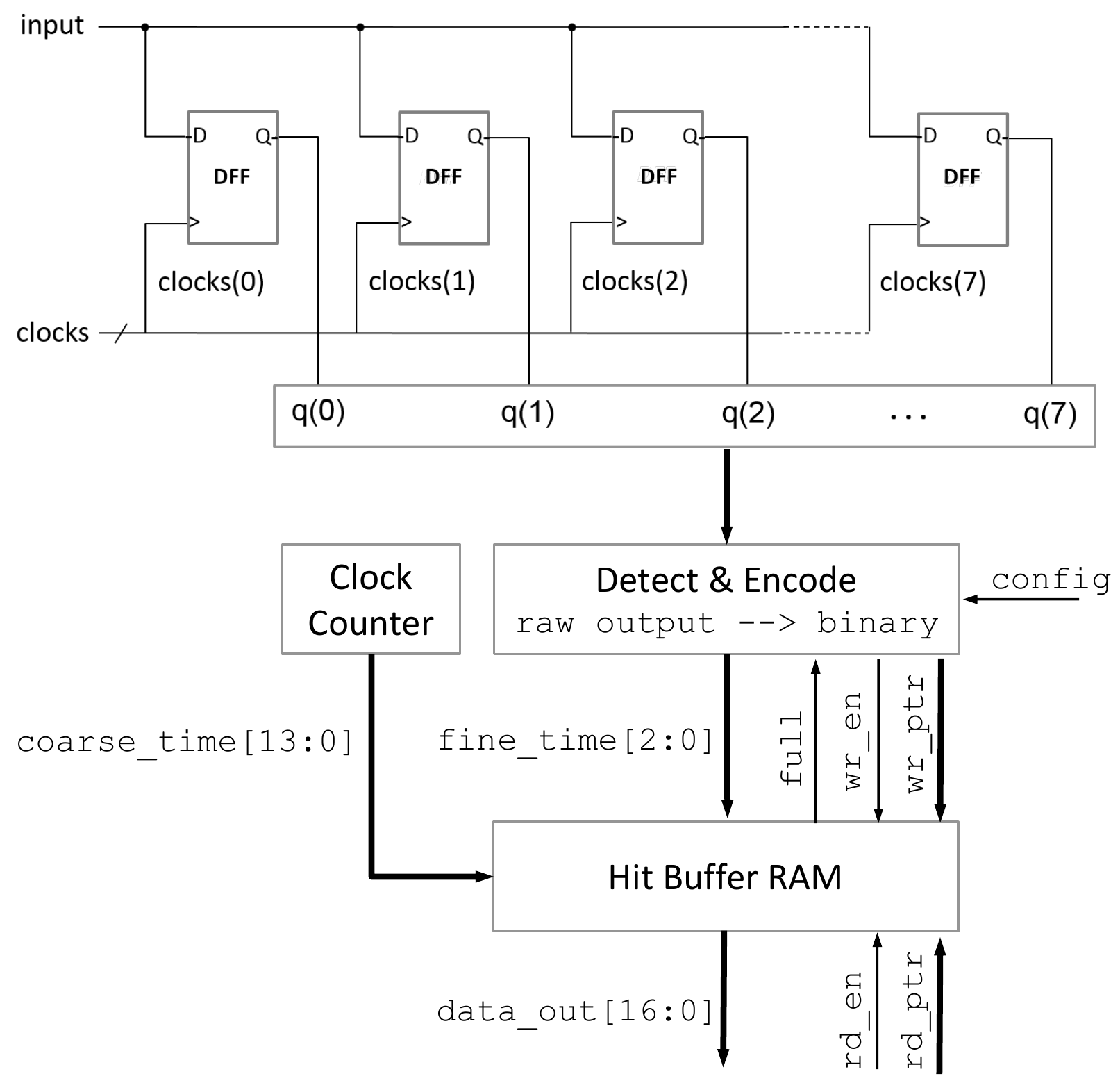}
        \caption{\label{fig:tdc_channel}}
    \end{subfigure} \hfill
    \begin{subfigure}[b]{0.57\textwidth}
        \includegraphics[width=\textwidth]{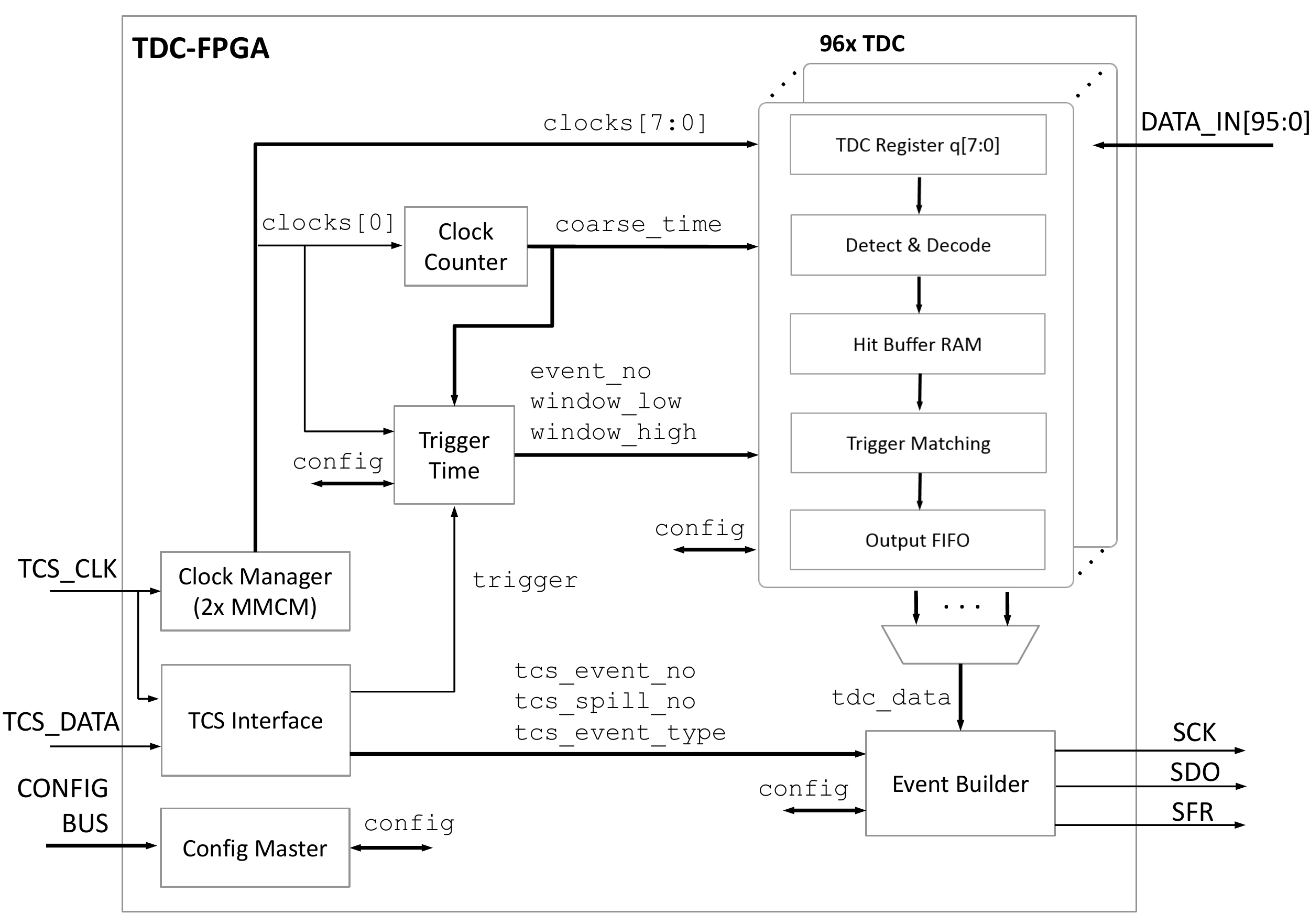}
        \caption{\label{fig:tdc_overview}}
    \end{subfigure}
    \caption{(\subref{fig:tdc_channel}) Block diagram of a TDC
      channel. (\subref{fig:tdc_overview}) Top-level view of the TDC-FPGA design.}
     \label{fig:tdc_design}
\end{figure}

The TDC application implements 96~channels on a single Artix-7 FPGA. The time digitization
of the incoming hits is accomplished by sampling the state of the input signal with a
quantization step (LSB) of 400\,ps. A coarse counter running at sampling clock frequency
extends the dynamic range to 16~bits. The result of the fine time measurement is stored
together with the coarse time tag in a (\mbox{2k~x~18}) integrated memory cell. The entire
digitization and readout process is dead time free. Figure~\ref{fig:tdc_design} shows the
block diagram of a TDC channel and the top-level view of the FPGA design to enhance the
comprehensibility of the technical description. The benchmarks of the TDC application are
summarized in table~\ref{tab:tdc_specs}.

\begin{table}[htbp]
    \centering
    \caption{\label{tab:tdc_specs} TDC-FPGA specifications.}
    \smallskip
    \begin{tabular}{|l|c|}
      \hline
      Number of channels & 96 \\
      \hline
      Time bin size (LSB) & 400\,ps \\
      \hline
      Sampling clock frequency & 311.04\,MHz \\
      \hline
      Average time resolution & 165\,ps \\
      \hline
      Differential non-linearity & <~0.3\,LSB\\
      \hline
      Integral non-linearity & <~0.3\,LSB\\
      \hline
      Input rate & Typical 15\,MHz \\
      \hline
      Double hit resolution & 3.2\,ns\\
      \hline
      Dynamic range & 16\,bits \\
      \hline
      Dead time & None \\
      \hline
    \end{tabular}
\end{table}

\subsection{Time digitization principle}
\label{sec:time-digit-princ}

Our method connects each input signal to a set of eight edge-triggered flip-flops. The
clock inputs of the associated register are driven by a multiphase clock with frequency
$f$. The phases are evenly aligned, activating the flip-flops in sequential order with a
delay step $\tau = 1/(8f)$. The sampling clock is synthesized from an external reference
clock input. The ARAGORN front-end was primarily designed to be operated at the COMPASS
experiment~\cite{compass2} at CERN. In that case, the sampling clock frequency
(cf.~table~\ref{tab:tdc_specs}) is adjusted to match an integer multiple of the COMPASS
reference clock. Two Mixed-Mode Clock Manager (MMCM) primitives inside the FPGA fabric are
used to control the phase alignment. The MMCM architecture is based on a phase-locked loop
(PLL) and thus automatically maintains a stable characteristic of the converter circuit
against supply voltage and ambient temperature variations.

The phase-shifted clocks are distributed on the global clock network of the FPGA,
especially designed for low skew and jitter tolerance, to achieve the best possible
uniformity of the transfer characteristic. Likewise, any skew present in the input signal
to the TDC register contributes to the non-linearity, but automatic routing with similar
propagation delay to all net endpoints is not supported. Nonetheless, an interactive
routing mode is provided with the Xilinx Vivado GUI to optimize the critical paths. The
optimal results obtained for a single TDC register were exported as constraints to lock
down the input signal routing in subsequent implementation runs. In order to fix the
routing of a net, the placement of the components involved has to be locked as well. A
script-based flow was used to apply the same low-skew routing in the implementation of all
channels. When the critical portions of the TDC input nets are properly constrained,
re-synthesis does not change the transfer characteristic, allowing for design updates
without larger effort.

After synchronization, the fine time portion of the measured timestamps is encoded from
the register output. The encoding logic can be configured during operation for leading,
trailing or even both edge sensitivity. The fine time corresponds to the bin number a hit
was found in the register. A coarse counter counting the periods of the sampling clock
delivers the coarse time to extend the measurement range. The measurement result combines
the fine time information and the coarse counter value. Encoded timestamps are stored in
dual-port hit buffers which allow for a dead time free digitization and readout process.

\subsection{Trigger matching}
\label{sec:trigger-matching}

In state-of-the-art high-energy physics experiments, pre-processing of digitized data as
early as on the front-end boards is essential. In our application, this is achieved with
an advanced trigger matching feature. This algorithm selects only such hits from the hit
buffers which are time-correlated to the trigger primitives. The time of trigger arrival
is digitized with coarse counter precision and a programmable latency time is subtracted
to account for the trigger generation and distribution delay. The corrected trigger time
defines the lower limit of the acceptance window. The upper limit is determined by a
configurable gate time that corresponds to the drift time or time of flight in the
detector. Timestamps matching the selective time window are considered to coincide with
the trigger event and copied to the output FIFOs. As the trigger matching process might be
busy while subsequent triggers arrive, the acceptance window borders are stored in a
trigger FIFO for each channel together with an identifier tag.

The search process scans through the hit buffer entries starting from a given memory
address until all matching hits are found or the read pointer catches up with the current
write address pointer. Finally, the read pointer is relocated to the first entry in the
acceptance window as older timestamps are no longer of interest and can be discarded. If
for longer periods the trigger matching is idle, the write pointer may catch up with the
search start address pointer at some point and assert the buffer full flag. In this case,
new hits would be lost. In order to speed up the operation and to prevent buffer overflow
conditions, artificial triggers are generated at regular intervals to update the search
start address pointer in the hit buffers. Hence another challenge of this algorithm is
that coarse counter rollovers invert the comparison operations of the timestamps with the
acceptance limits. Thus, the timestamps stored in the hit buffers include an additional
rollover bit that is omitted during transfer to the output FIFOs.

\subsection{Readout interface}
\label{sec:readout-interface}

After trigger matching, the corresponding data sets stored in the individual output FIFOs
have to be concentrated into a single data stream prior to data output. This is done in a
multistage scheme reading the output FIFOs in parallel and buffering the event data for
maximum data throughput. In the final stage, the associated event labels are added and the
data packages are transfered via a high-speed serial link to the central FPGA. The
employed data concentrator modules are highly configurable and easily cascadable to
enhance the maintainability of the design.

\subsection{Measurement results}
\label{sec:measurement-results}

\begin{figure}[htbp] \centering 
 \begin{subfigure}[b]{0.49\textwidth}
        \includegraphics[width=\textwidth]{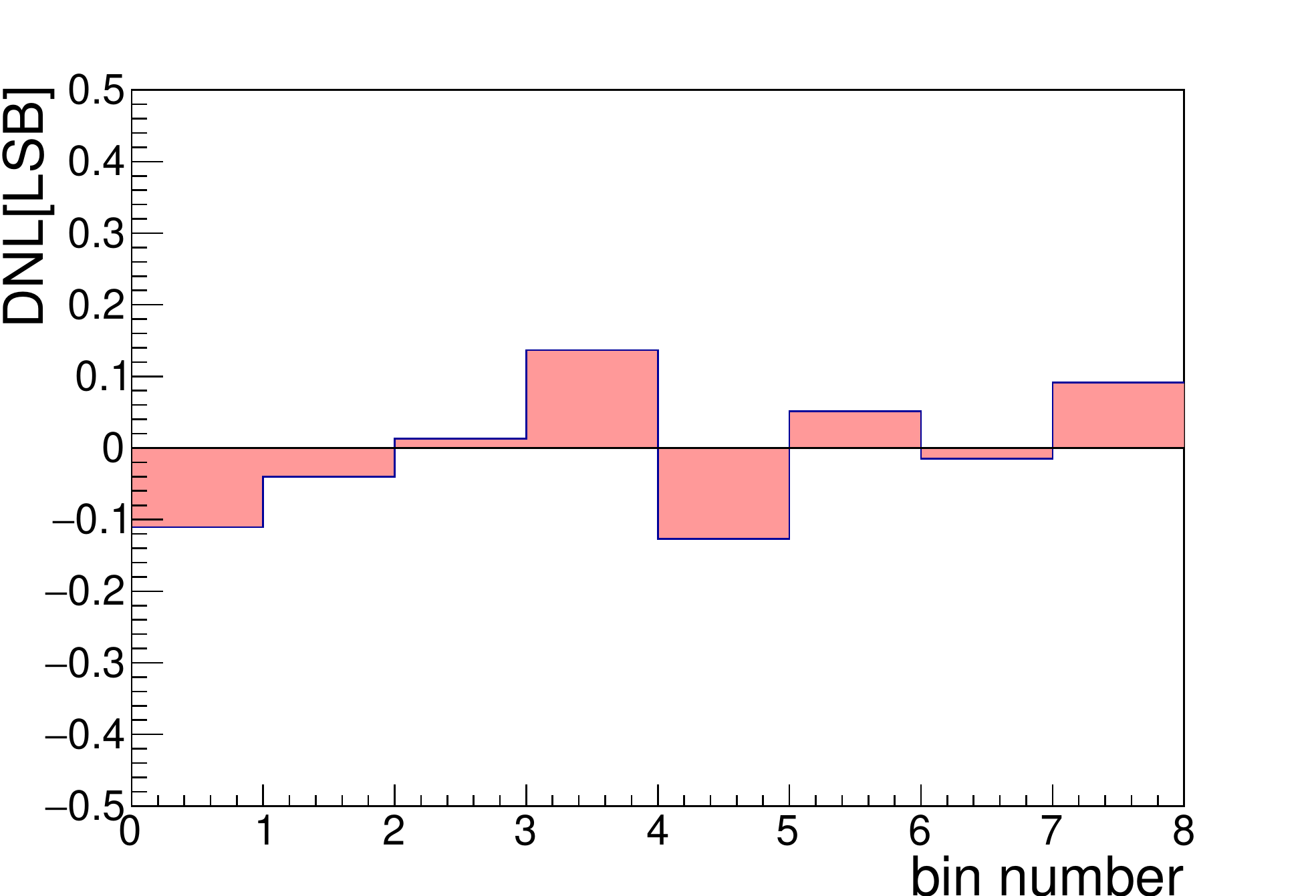}
        \caption{\label{fig:dnl_min}}
    \end{subfigure} \hfill
    \begin{subfigure}[b]{0.49\textwidth}
        \includegraphics[width=\textwidth]{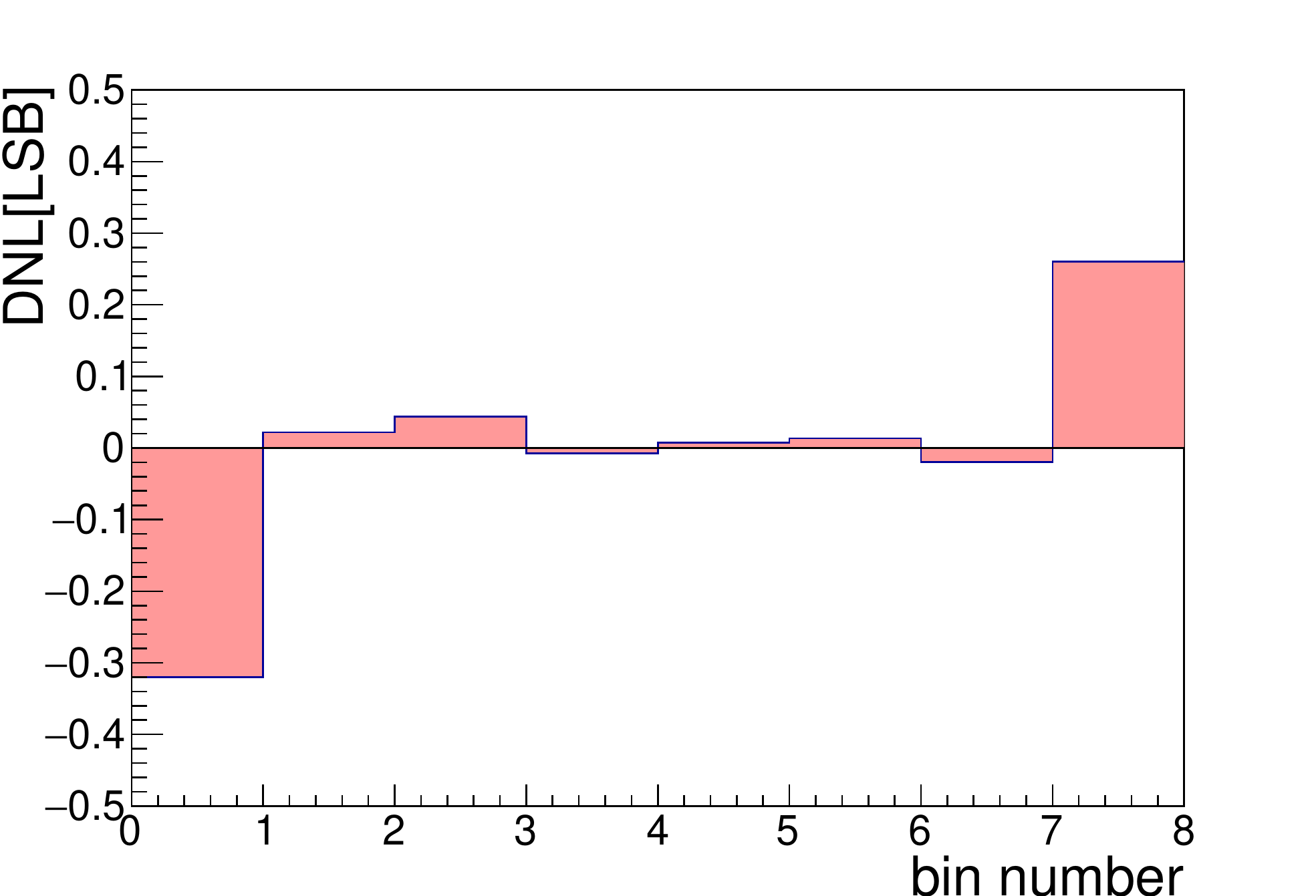}
        \caption{\label{fig:dnl_max}}
    \end{subfigure}
    \caption{Channel with minimum (\subref{fig:dnl_min}) and maximum
      (\subref{fig:dnl_max}) DNL.}
     \label{fig:dnls}
\end{figure}

The transfer characteristic of the TDC was measured with a statistical code density test. The
test signals originated randomly in time from a function generator. Fanout buffer extension
boards were employed to cover all inputs of the ARAGORN front-end. If the input signals are
uncorrelated with the sampling clock, each time bin of an ideal TDC receives the same number of
entries  ${n=N/8}$. Assuming a sufficiently large number of measurements $N$, the
statistical error $1/\sqrt{n}$ can be neglected. Due to the imperfections described in
section~\ref{sec:time-digit-princ}, a differential non-linearity (DNL) is obtained in every bin

\begin{align}
  \label{eq:dnl}
  DNL_i  &= \frac{n_i-n}{n}.
           \intertext{Figure~\ref{fig:dnls} shows the result for the channel with minimum and maximum
           DNL. The
           standard deviation $\sigma$ of a measured time interval $\Delta t$ strongly depends on the
           fractional part $f$ of the quotient $\Delta t/LSB$~\cite{kalisz}}
  \label{eq:sigma}
  \sigma(f) & = LSB \sqrt{(1-f)f}.
\intertext{The average standard deviation $\sigma_{avg}$ is determined by integration of
              (\ref{eq:sigma}) within the limits ${0<f<1}$}
\label{eq:sigma-avg}
\sigma_{avg} &= \frac{\pi}{8}LSB \approx 0.39 LSB.
\end{align}

\begin{figure}[htbp]
    \centering 
    \includegraphics[trim={0 0 0 2mm},clip,width=.5\textwidth]{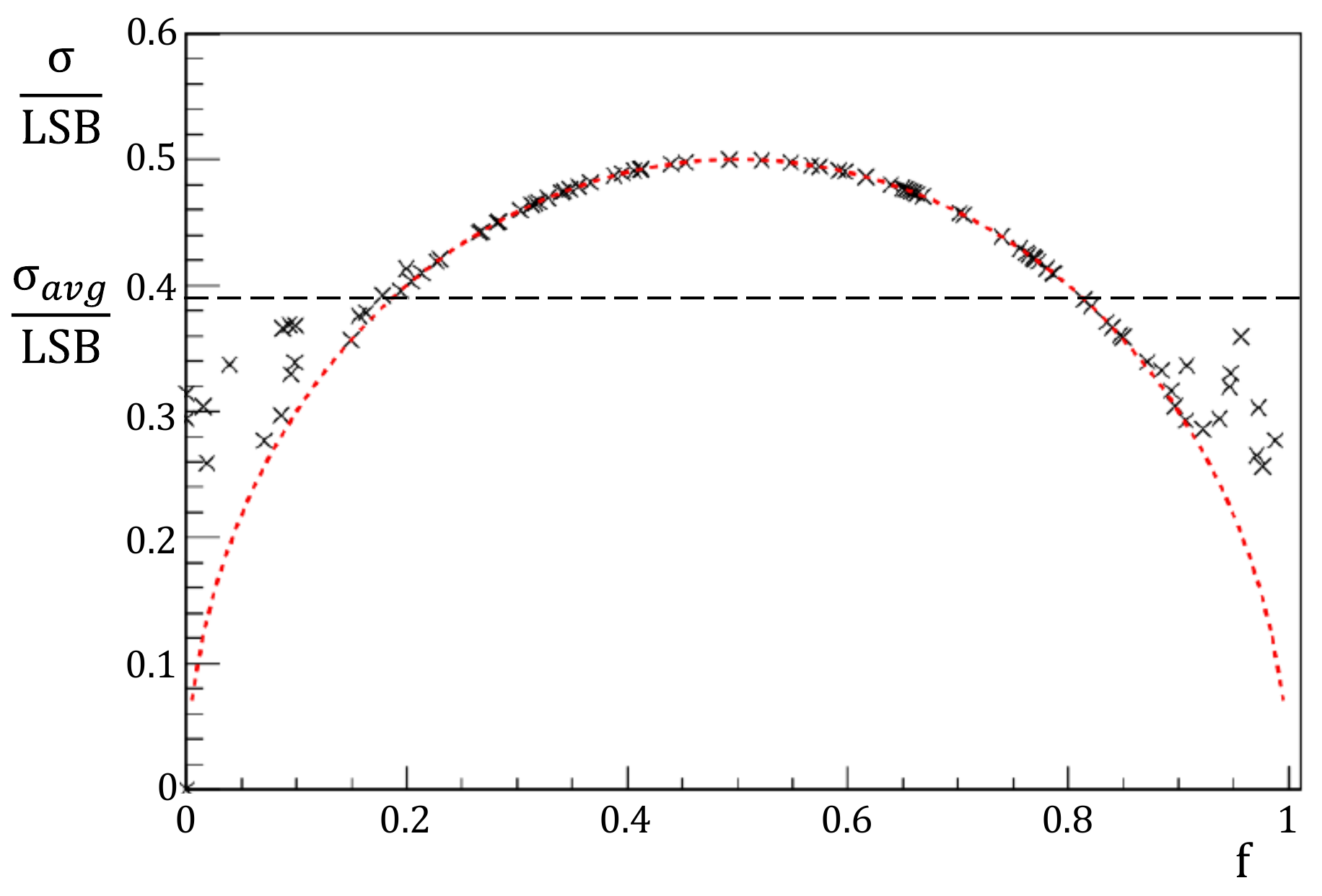}
    \caption{\label{fig:resolution} Time resolution by fractional part $f$ of the quotient
      $\Delta t/LSB$. The ideal case (\ref{eq:sigma}) and the average time resolution
      (\ref{eq:sigma-avg}) are shown as dashed lines.}
\end{figure}

Determining the time resolution of the TDC, we measured constant time intervals between
two channels. The calculated standard deviations for all channel of a TDC-FPGA were
plotted against the fractional part of the time interval mean divided by LSB. As
propagation delays of the input signals on the PCB vary, figure~\ref{fig:resolution} shows
a uniform coverage in the interval $0<f<1$. The results are in good agreement with the
half circle curve corresponding to (\ref{eq:sigma}) apart from the outer regions where the
non-linearity of the transfer characteristic takes effect. The average time resolution of
all channels calculated on the basis of this measurement is 165\,ps. For the ideal TDC,
the time resolution given by (\ref{eq:sigma-avg}) is 157\,ps.

\section{Constant latency link}
\label{sec:const-latency-link}

In the uplink direction, the optical network distributes the trigger primitives and the
reference clock employed in the TDC application to the front-end boards. This requires
that the reference clock recovered in the transceiver tiles maintains a constant phase
relationship on all front-ends after a power cycle, reset or a loss of lock. Similarly,
the trigger and miscellaneous control signals are expected to approach the receiver nodes
with predictable latency, e.g. to simultaneously reset all coarse counter primitives at
system startup. These requirements demand for a custom configuration because some features
of the GTP high-speed transceivers~\cite{ug_gtp} embedded in the Artix-7 FPGA fabric show
latency variations.

Within the scope of high-energy physics experiments, the benefit of synchronous trigger
and clock distribution systems based on FPGA transceiver links have been demonstrated and
discussed for some time, e.g. in~\cite{giordano}. Besides, our design requires synchronous
retransmission of the receiver link data from the master front-end to the slave boards as
outlined in section~\ref{sec:intro}. To comply with the strict jitter specifications of
the transceiver reference clock source, the ARAGORN front-end features a jitter
attenuator. We selected the Texas Instruments LMK04906 device that meets the above
prerequisites. The device reliably removes jitter from the recovered clock providing the
transceiver tiles and as well the TDC-FPGAs with a clean reference clock.

The GTP transceivers are organized in groups of four, the transceiver quads, sharing two
differential external reference clock input pin pairs and two PLLs for the high-speed
serial transmitter and as clock seed for the CDR\footnote{Clock Data Recovery}
circuit. Every GTP transmitter/receiver channel consists of a PMA\footnote{Physical Medium
  Attachment} and a PCS\footnote{Physical Coding Sublayer} block. The PMA comprises the
serializer/deserializer logic, whereas the PCS handles the parallel data transfer from/to
the FPGA fabric. In the receiver PMA, the high-speed CDR clock output is divided to
produce the parallel recovered clock. This implies a variable phase offset between these
clocks and consequently latency variations in the data path. 

The alignment of the incoming data stream to the correct word boundaries is performed
internally in the PCS or by user intervention from the FPGA fabric. In the latter, the
dedicated RXSLIDE signal is asserted to shift the parallel data by one bit. If the total
number of bit shifts required for alignment is even, the bit-slip operation is performed
by shifting the phase of the parallel PMA clock only. For odd numbers, a bit shift in the
parallel data is involved and the clock phase differs. This is in fact the major reason
why a deterministic clock phase cannot be obtained with the standard transceiver
configuration. 

The solution was to bypass the data alignment and the 8b/10b decoder in the transceiver
and implement them inside the FPGA fabric. The parallel data bus of the FPGA interface is
20-bit wide. Our comma detection logic resets the GTP receiver until the alignment
sequence (K28.1 + K28.5) is found. This approach completes very quickly as in the link
idle state every fourth transmitted symbol is a comma character. The default reset
procedure was customized to include the jitter attenuator locked condition. Once the
uplink connection is established, the reset of the transmitter channels linked to the
slave boards is deasserted. It has become evident that the GTP phase adjust FIFO
introduces latency variations in the data path as well. Since the parallel fabric clock
and the reference clock originate from redundant outputs of the jitter attenuator, the
integrated phase align circuit could be exploited to resolve the phase differences between
the PMA clock domains, instead.

\begin{figure}[htbp]
    \centering 
    \includegraphics[width=.9\textwidth]{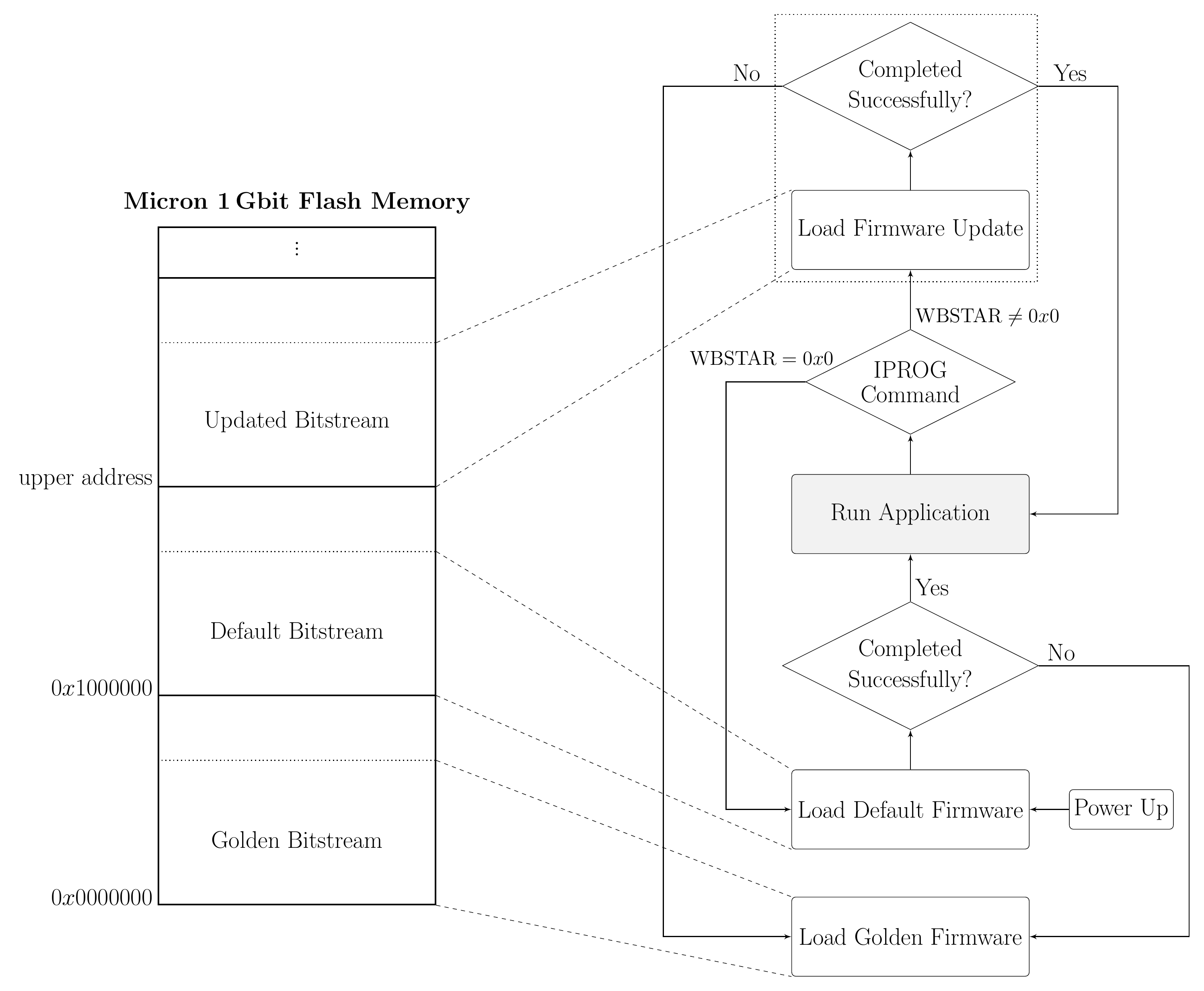}
    \caption{\label{fig:reconfigure} Reconfiguration flow chart.}
\end{figure}

\section{Embedded design}
\label{sec:embedded-design}

After power-up, the central FPGA retrieves the configuration bitstream from an onboard
Flash memory. When the constant latency link is established, the TDC-FPGAs are configured
with a bitstream delivered from remote storage. The central FPGA design implements an
embedded Microblaze processor to supervise the FPGA configuration, low-speed bus
interfaces and in-system Flash programming. 

A main objective of applications in environments with limited access like high-energy
physics experiments is to maintain remote communication with the readout systems at all
times. Figure~\ref{fig:reconfigure} illustrates our elaborated reconfiguration
strategy. If configuration with the default bitstream fails, e.g. due to a CRC error, a
fallback sequence is triggered to load a 'golden bitstream' from the Flash base
address. The golden bitstream allows the user to fix up corrupted Flash content and to
reconfigure from a given address space. Loading multiple images on the fly is achieved by
sending a dedicated command to the Internal Configuration Access Port (ICAPE2)
primitive. This feature is highly beneficial for bitstream upgrades and debugging in the
field.

\section{Conclusion}
\label{sec:conclusion}

In conclusion, we present a highly versatile TDC platform with outstanding high-speed
optical readout capabilities. This novel approach permits 3072 input channels to be
concentrated and read out with a single optical fiber at a bandwidth of 6.6\,Gb/s. Another
highlight is the superior channel density of a single front-end with a form factor of
\mbox{140\,mm~x~172\,mm} comprising 384 TDC inputs, limited only by connector spacing
constraints.  Accordingly, we succeeded in the development of a very cost-optimized
design. A time resolution of 165\,ps for all channels was obtained for the TDC
application. The design consumes only 16\% of the flip-flop registers and 22\% of the
look-up tables (LUTs) available in the Artix-7 FPGA. This will allow for future upgrades
if applications demand for even higher precision. The digitization bin size could be
easily divided in half by doubling the length of the TDC register and engaging the falling
edges of the multiphase sampling clock. We furthermore demonstrated the implementation of
a constant latency link with the embedded high-speed GTP transceivers. Our approach
provides fixed latency data transfer and deterministic phase-synchronous clock
distribution even in multi-tiered front-end arrangements. In-system reconfiguration and
firmware upgrade strategies have also been developed to entirely oversee and maintain the
readout chain remotely. In addition, a system-wide bus interface allows for slow control
tasks ranging from register access with command line tools to complete configuration file
upload.

\acknowledgments
The authors gratefully acknowledge the support of the local electronic workshop during the
front-end design phase. We also appreciate the fruitful discussions with our colleagues
from the COMPASS front-end and RICH groups and the endeavours of our collaborators during
several testing periods at CERN. This work is supported by the Bundesministerium für
Bildung und Forschung (BMBF) and EU FP7 (Grant Agreement 283286).

\end{document}